# Interplay between Spinodal Decomposition and Glass Formation in Proteins Exhibiting Short-Range Attractions

Frédéric Cardinaux,* Thomas Gibaud, Anna Stradner, and Peter Schurtenberger[†]

*Department of Physics, University of Fribourg, CH-1700 Fribourg, Switzerland*
(Received 16 December 2006; published 13 September 2007)

We investigate the competition between spinodal decomposition and dynamical arrest using aqueous solutions of the globular protein lysozyme as a model system for colloids with short-range attractions. We show that quenches below a temperature $T_a$ lead to gel formation as a result of a local arrest of the protein-dense phase during spinodal decomposition. The rheological properties of these gels allow us to use centrifugation experiments to determine the local densities of both phases and to precisely locate the gel boundary and the attractive glass line close to and within the unstable region of the phase diagram.

Colloidal suspensions have frequently been used as ideal model systems to address fundamental issues in condensed matter physics such as liquid ordering, crystallization, and glass formation [1]. Particular attention has been given to fluid-solid transitions, where for some appropriate choice of experimental conditions particles arrest and form disordered solids. This phenomenon has been investigated in the two limiting cases of hard spheres [2] and strongly attractive particles [3,4]. However, while these two classes of model systems have been thoroughly studied both experimentally as well as theoretically, they have mostly been treated using completely different approaches and view points. Mode coupling theory (MCT) has been used to interpret the hard sphere glass transition [5], while diffusion limited cluster aggregation models successfully describe the formation of irreversibly aggregated fractal gels in the limit of low volume fractions $\phi$ and very deep attractive potential wells [6].

However, recent experiments, theory, and computer simulations have revealed striking analogies [7–10] between colloidal glasses and gels and have stimulated an increased effort to unify the description of the transitions to these disordered solidlike states within a single conceptual framework. A generic state diagram has emerged from this effort in characterizing dynamical arrest in attractive particle suspensions [6,11,12], but many questions remain unanswered and a theoretical picture unifying these two limits is clearly still missing. This is particularly true for the case of intermediate $\phi$ and a strength of the attraction of $\lesssim 10 k_B T$, where the system may undergo a liquid-gas phase separation that intervenes with dynamical arrest.

Early pioneering work of Vrij, Dhont, and collaborators revealed evidence for an arrested spinodal decomposition of attractive colloidal particles, where complete macroscopic phase separation is hindered by an arrest transition of the dense phase at $\phi$ much smaller than those estimated from the high density branch of the equilibrium coexistence curve [13,14]. Subsequently, the existence of "transient gels" was described in colloid-polymer mixtures [15] where fingerprints of spinodal decomposition were identified in the measured static structure factor [16]. Percolation [13,16] as well as an attractive glass transition [17–19] have been proposed to account for this local arrest. The latter is supported by a recent study where MCT predictions using experimental structure factors have been mapped to measure nonergodicity factors with the local $\phi$ of arrest as a free parameter [20]. However, for intermediate and strong interparticle attractions there still exists the unresolved issue [21] whether dynamical arrest can only occur via an "arrested" spinodal decomposition, or whether there exists also an "equilibrium" route to gelation where the gel line can become more stable than the coexistence curve and exist above and to the left of the binodal [18,22]. There is thus a clear need for systematic experimental data, in particular, with respect to the exact location of the glass or gel line close to and within the coexistence curve for different interaction potentials [21].

In this Letter we describe the use of the globular protein lysozyme as a convenient model system to investigate this interplay between spinodal decomposition and dynamical arrest in colloids with short-range attractions. We find clear evidence for a scenario of dynamical arrest where a sufficiently deep temperature quench below the spinodal first leads to spinodal decomposition that then becomes arrested once the dense phase forms an attractive glassy state. Based on the specific rheological properties of these arrested samples, we are then able to use centrifugation experiments to separate the glassy phase and quantitatively determine the location of the arrest line in the unstable region of the phase diagram.

We use hen egg white lysozyme (Fluka, L7651), a 14.4 kDa protein, in 20 m$M$ Hepes buffer at $p$H = 7.8 containing 0.5$M$ NaCl. Initially a stock solution at $\phi \approx$ 0.22 is prepared in pure buffer without added salt, and its $p$H is adjusted to 7.8 ± 0.1 with NaOH [23,24]. We then dilute it with a NaCl-containing buffer to a final NaCl concentration of 0.5$M$. Particular care is taken to avoid partial phase separation upon mixing by preheating both buffer and stock solution well above the liquid-liquid coexistence curve. This procedure results in completely transparent samples at room temperature with $\phi$ ranging from 0.01 to 0.18, where $\phi$ was obtained from the protein concentrations $c$ measured by UV absorption spectroscopy using $\phi = c/\rho$, where $\rho = 1.351$ g/cm$^3$ is the protein density. To prepare samples at high $\phi$ up to 0.34, we use phase separation into a protein-rich and protein-poor phase. Typically, a sample at $\phi = 0.155$ is quenched to a temperature $15\,°C < T < 18\,°C$ below its cloud point and centrifuged at 9000 g for 10 minutes. Quasiequilibrium is reached once the two phases are separated by a sharp meniscus and show only slight turbidity. The bottom dense phase is used for further experiments.

In a first step we determine the coexistence curve for liquid-liquid phase separation over a $\phi$ range of 0.01 to 0.34 using cloud point measurements. Solutions of known $\phi$ in 5 mm NMR tubes are placed in a $T$-controlled bath well above the binodal, $T$ is slowly decreased, and the cloud point is obtained at the temperature at which samples become turbid. Additionally, we perform dynamic and static light scattering measurements at 90° using a 3D-LS setup (LS-Instruments GmbH, $\lambda = 633.8$ nm) that allows for efficient suppression of multiple scattering. An extrapolation of the inverse of the single scattered intensity as a function of $T$ then yields an upper limit for the spinodal temperature [25]. It is worth mentioning that the equilibrium state of all studied state points is a liquid coexisting with crystals [26], as expected for short-range attractive systems where the range of the attraction is $\lesssim 0.25\%$ of the particle diameter [15,21], and our experiments were performed before the onset of crystallization. The resulting coexistence curves for liquid-liquid phase separation are shown in Fig. 1(a), with the corresponding binodal (open circles) and spinodal (open squares) in full agreement with previous studies on lysozyme [26].

We then turn to a closer examination of the samples quenched into the spinodal region, where we systematically vary $T$ of the final state. The set of phase contrast micrographs presented in Fig. 1(b) illustrates the typical demixing kinetics for two different quenches. For a shallow quench to $T = 16.8\,°C$ (∗), we observe the classical sequence of domain formation and coarsening in spinodal decomposition that ultimately lead to complete demixing to $\phi$ given by the coexistence curve. However, for quenches below $15\,°C$ (×) phase separation follows a completely different route. Here, the spinodal domain structure initially coarsens but then completely arrests after typically 30 s, without any signs of structural evolution during the observation period of 8 hours. Moreover, we find that this arrest occurs at a temperature $T_a = 15 \pm 0.3\,°C$ that is remarkably constant independent of the initial $\phi$. This arrest of the domain structure appears in the region delimited by the spinodal at low $\phi$ and by the partial tie line at $T_a$ (dashed line in Fig. 1).

These observations are in agreement with a mechanism, where a solidlike gel is formed whose connectivity is provided by the bicontinuous nature of the spinodal decomposition process while its rigidity arises due to a glass-like arrest of the dense phase as its composition crosses the glass line [17,20]. The arrest temperature $T_a$ thus corresponds to the intersection point of the attractive glass line with the coexistence curve at $\phi_a = 0.34$. To obtain the position of the attractive glass line at lower $T$ we use a centrifugation method which can be applied because of the specific rheological properties of the arrested samples at $T < T_a$. We use a stress-controlled rheometer (MCR300 from Paar Physica) with a cone and plate geometry and a solvent trap to minimize sample evaporation. The samples are loaded in the rheometer at $T$ well above the binodal and subsequently quenched to the final $T$. The phase separation is allowed to proceed in the rheometer for a period of 300 s before performing two interleaved creep tests with an oscillatory measurement in between. The creep tests, where the strain $\gamma(t)$ following a step stress $\sigma$ with 0.2 Pa $< \sigma <$ 2 Pa (for the data of Fig. 2(b) $\sigma = 2$ Pa) is recorded up to 1200 s, give us access to the long-time response of the relaxation spectrum. The short-time response is provided by the oscillatory measurements probing frequencies from 0.01 to 100 Hz. We carefully checked that the applied stress, respectively, strain amplitudes, were always low enough to ensure linear response.

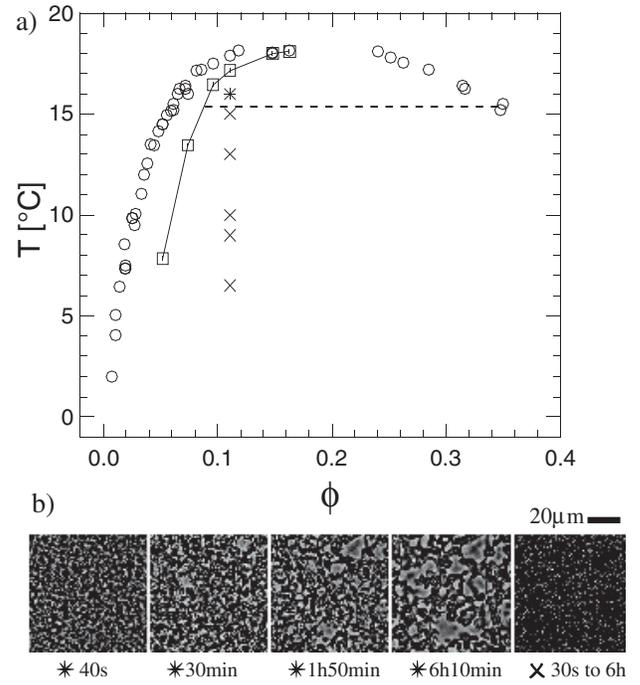

FIG. 1. (a) $T$-$\phi$ plane of the phase diagram of aqueous lysozyme solutions (20 m$M$ Hepes buffer, $p$H = 7.8, 0.5$M$ NaCl). Liquid-liquid coexistence curve (◯), spinodal (□). Also shown are state points in the unstable region investigated with rheology where liquidlike (∗) and solidlike (×) behavior has been observed. (b) Phase contrast micrographs of samples at $\phi = 0.11$ showing the coarsening at 16.8 °C (∗) and the freezing at 13 °C (×) of the bicontinuous texture in the spinodal region.

A representative result of the oscillatory measurements on a sample at $\phi = 0.148$ quenched to $T = 13\,°C < T_a$ is presented in Fig. 2(a). We find at high frequencies a solid-like response characterized by an elastic modulus $G'(\nu)$ larger than its loss modulus $G''(\nu)$. At low frequencies a dissipative regime is revealed by the crossover of $G''(\nu)$ with $G'(\nu)$. The access to longer relaxation times is provided by the creep test shown in Fig. 2(b). The dissipative regime is now visible at very short times through the initial decay of the inverse creep compliance $J^{-1}(t) = \sigma/\gamma(t)$. At longer times, $J^{-1}(t)$ levels off and becomes constant at a value that can be associated to the long-time elastic modulus $G_\infty \propto J^{-1}(\infty)$. The existence of $G_\infty$ confirms the solid-like nature of the sample. It is important to point out that this feature is obtained for all quenches into the gel region, whereas for $T > T_a$ in the spinodal region we find a liquid-like response with $J^{-1}$ fully decaying to zero. We also note that the two successive creep tests gave identical results, demonstrating that the structure of the gels is neither broken by the applied stress nor affected by further coarsening over the measurement period of typically 1 h.

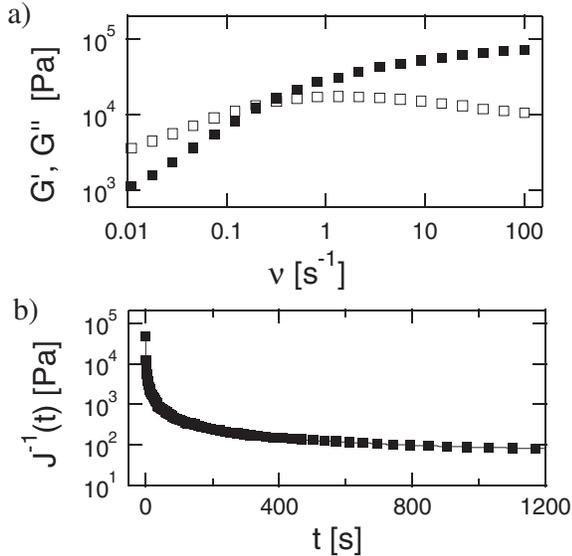

FIG. 2. (a) Elastic modulus $G'(\nu)$ (full symbols) and loss modulus $G''(\nu)$ (open symbols) for a gel formed in the spinodal region at $\phi = 0.148$ and a quench temperature $T = 13\,°C < T_a$. (b) Corresponding inverse creep compliances $J^{-1}(t)$.

The rheological properties of these gels can be understood at least qualitatively based on the proposed mechanism of an arrested spinodal decomposition. Two characteristic length scales directly emerge from this scenario [20]: the mesh size $\xi$ of the spinodal structure at the point of arrest and the characteristic length found in dense attractive glasses as the underlying basis for gel formation. Fast Fourier transforms of micrographs from arrested gels yield an estimate of $\xi \sim 1\,\mu m$, whereas we expect a characteristic length of the order of the particle diameter (3.4 nm) for the protein-dense phase where $\phi$ can be as high as $\phi_a$. This large size difference provides an explanation for the well-separated temporal responses observed experimentally. At high frequencies, the oscillatory measurements mainly probe the local mechanical response of the glassy dense phase, whereas $G_\infty$ reflects the additional contributions from the gel superstructure with the much larger mesh size. At intermediate times, the different relaxation modes of the strands forming the network lead to the observed dissipative regime.

The large difference in the observed elastic plateaus, typically $G'(100\,Hz)/G_\infty \gtrsim 10^2$, also suggests that our system possesses two very different yield stresses. This allows us to further investigate the location of the arrest line in the unstable region of the phase diagram using centrifugation. At sufficiently high centrifugal force fields, we expect the large gel structures to yield without inducing a fluidization of the glassy dense phase. The centrifugation will then lead to a macroscopic separation of the previously entrapped low-density protein fluid and the high density protein glass. We observe the formation of a sharp interface between a liquid phase on top and a homogeneous glass phase with a density that provides the position of the glass line at this particular temperature. This is illustrated in Fig. 3(a), where we monitor the interface height $h$ with time for samples with $\phi = 0.155$ quenched and centrifuged at 1080 g at different $T < T_a$.

For all quench depths, $h$ decreases rapidly and reaches a constant plateau value $h_{pl}$ after 30 minutes. Moreover, the fact that the concentration of the upper low-density fluid phase remains constant at all centrifugation times clearly indicates that the centrifugation does not affect the phase coexistence but instead breaks the spinodal network and sediments the arrested regions only. To ensure that no further compaction of the solid phase occurs, we perform a systematic variation of the acceleration value and measure the resulting values of $h_{pl}$. Figure 3(b) shows that for all quench depths there exists a region where $h_{pl}$ does not depend on the applied $g$ values.

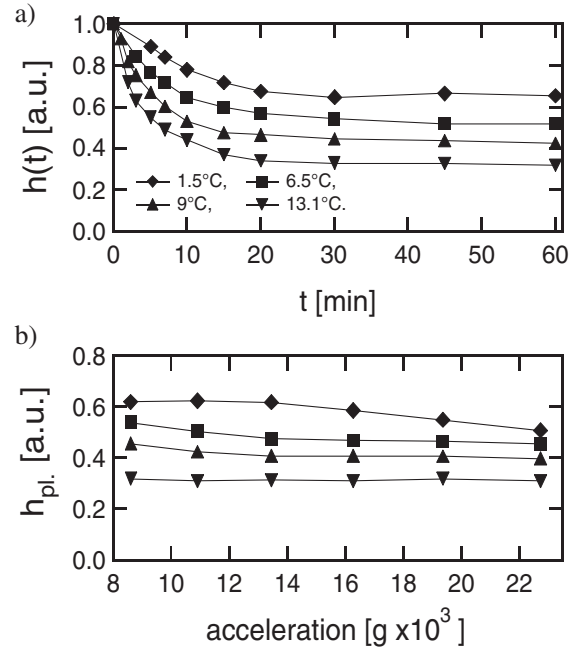

FIG. 3. (a) Evolution of the relative height of the interface, $h(t)$, with centrifugation time for different quenched temperatures. (b) Dependence of the plateau value $h_{pl}$ relative to imposed acceleration. Same symbols as in (a).

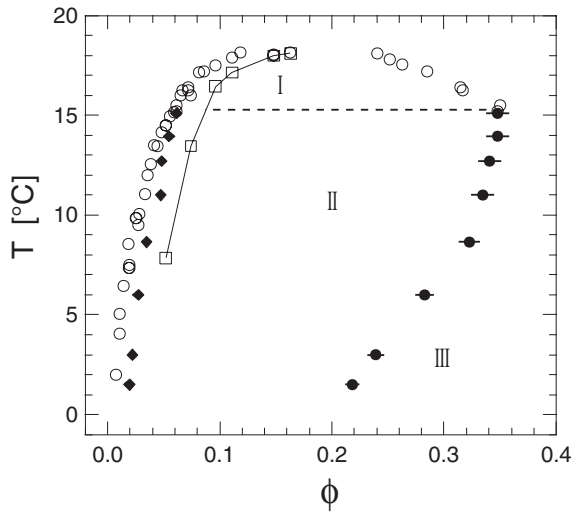

FIG. 4. Kinetic phase diagram of aqueous lysozyme solutions showing regions of complete demixing (I), gel formation (II), and glasses (III). Full symbols stand for the results of the centrifugation experiments: (●) arrested dense phase, (◆) dilute phase.

The resulting $\phi$ for the dilute and dense phases are reported as the full symbols in Fig. 4. The local arrest line formed by the volume fractions of the glassy phase (●) crosses the coexistence curve at $\phi_a$ and $T_a$, in agreement with the previously determined intersection point, and it extends deep into the unstable region. It delimits a region where homogeneous attractive glasses can be reached by quenches at sufficiently high $\phi$ and low $T$ (III in Fig. 4) from a large region (II) where gels are formed via an arrested spinodal decomposition. Microscopically, these gels correspond to a coexistence of a dilute fluid with a dense percolated glass phase. Finally, shallow quenches into region I lead to complete demixing. It is interesting to note that the protein concentrations in the low-density phase (◆) are higher than those predicted by the equilibrium coexistence curve, reflecting the fact that the phase coexistence and the resulting concentrations of the two phases have been altered by the arrest of the spinodal decomposition [19].

The use of proteins as model colloids interacting via a short-ranged attraction has allowed us to investigate the interplay between phase separation and dynamical arrest in colloidal suspensions. Our experiments directly support recent computer simulations of short-ranged attractive particles [17,18]. There it has been concluded that for this class of potentials disordered arrested states at low $\phi$ can only be created under out-of-equilibrium conditions, requiring a preliminary gas-liquid separation into colloid-rich and -poor phases followed by glasslike arrest in the denser regions. For the first time, we are now able to quantitatively locate the attractive glass line in the unstable region below the spinodal. The kinetic phase diagram of Fig. 4 provides a new test ground for computer simulations and theoretical calculations in the current attempt to understand and generalize dynamical arrest in soft matter.

We are deeply grateful for fruitful discussions with Veronique Trappe. This work was supported by the Swiss National Science Foundation, the State Secretariat for Education and Research (SER) of Switzerland, and the Marie Curie Network on Dynamical Arrest of Soft Matter and Colloids (No. MRTN-CT-2003-504712).

---

*Current address: Physik der weichen Materie, IPkM, Heinrich-Heine-Universität Düsseldorf, 40225 Düsseldorf, Germany.
†Corresponding author.